\documentclass[aps,prl,twocolumn,superscriptaddress,groupedaddress]{revtex4}  
\usepackage{graphicx}  
\usepackage{dcolumn}   
\usepackage{bm}        
\usepackage{amssymb}   

\begin{document}

\title{Tunneling spectroscopy of graphene nanodevices coupled to large-gap superconductors}

\author{Joel I-Jan Wang}
\affiliation{Department of Physics, Massachusetts Institute of Technology, 77 Massachusetts Avenue, Cambridge, Massachusetts 02139, United States}

\author{Landry Bretheau}
\thanks{landry.bretheau@polytechnique.edu}
\affiliation{Department of Physics, Massachusetts Institute of Technology, 77 Massachusetts Avenue, Cambridge, Massachusetts 02139, United States}
\affiliation{Laboratoire des Solides Irradi\'es, \'Ecole Polytechnique,
CNRS, CEA, 91128 Palaiseau, France}

\author{Daniel Rodan-Legrain}
\affiliation{Department of Physics, Massachusetts Institute of Technology, 77 Massachusetts Avenue, Cambridge, Massachusetts 02139, United States}

\author{Riccardo Pisoni}
\affiliation{Solid State Physics Laboratory, ETH Z{\"u}rich, CH-8093 Z{\"u}rich, Switzerland}

\author{Kenji Watanabe}
\affiliation{National Institute for Materials Science, Namiki 1-1, Tsukuba, Ibaraki 305-0044, Japan}

\author{Takashi Taniguchi}
\affiliation{National Institute for Materials Science, Namiki 1-1, Tsukuba, Ibaraki 305-0044, Japan}

\author{Pablo Jarillo-Herrero}

\affiliation{Department of Physics, Massachusetts Institute of Technology, 77 Massachusetts Avenue, Cambridge, Massachusetts 02139, United States}

\date{\today}

\begin{abstract}
We performed tunneling spectroscopy measurements of graphene coupled to niobium/niobium-nitride superconducting electrodes.
Due to the proximity effect, the graphene density of states depends on the phase difference between the superconductors and exhibits a hard induced gap at zero phase, consistent with a continuum of Andreev bound states.
At energies larger than the superconducting gap, we observed phase-dependent energy levels displaying the Coulomb blockade effect, which are interpreted as arising from spurious quantum dots, presumably embedded in the heterostructures and coupled to the proximitized graphene.
\end{abstract}


\maketitle

The superconducting proximity effect in graphene has attracted considerable experimental interest this last decade, either via supercurrent measurements of graphene-based Josephson junctions
\cite{Heersche2007,Du2008,Girit2009,Ojeda-Aristizabal2009,Komatsu2012,BenShalom2016,Calado2015,Allen2016,Amet2016}, via tunneling spectroscopy 
studies of proximitized graphene ~\cite{Tonnoir2013,Natterer2016,Bretheau2017} or via 
phase transition measurements of tin-decorated graphene~\cite{Allain2012,Han2014}.
Moreover, graphene's extended two-dimensional nature
makes it a promising platform to explore the interplay of superconductivity with the quantum Hall effect, which could lead to the detection of exotic quasiparticles with nontrivial braiding statistics
\cite{Lindner2012,Clarke2013,Mong2014,San-Jose2015}.
To do so, it is necessary to strongly couple low-disorder graphene to large critical field superconductors.
Along this line, improvements in nanofabrication have led recently to the demonstration of high-field Josephson effect in ballistic graphene coupled to niobium, evidenced by Fabry-Perot oscillations of the supercurrent and anomalous Fraunhofer patterns \cite{BenShalom2016}. Even more recently, it was shown that the Josephson effect could persist in the quantum Hall regime by coupling a graphene sheet to molybdenum-rhenium~\cite{Amet2016}. Further studies are  however needed to elucidate the origin of these phenomena. Phase-controlled tunneling spectroscopy seems promising  as it enables one to probe Josephson physics in the energy domain~\cite{Bretheau2017}.

Microscopically, the Josephson effect arises from the formation in the normal conductor of entangled electron-hole states called Andreev bound states (ABS)~\cite{Kulik1970,Furusaki1991,Beenakker1991,Bagwell1992}. 
These fermionic states have energies inside the superconducting gap $[-\Delta,\Delta]$ that depend on the phase difference $\varphi$ between the two superconductors sandwiching the central conductor. Although phase-dependent spectroscopy of ABS has been performed in a few systems~\cite{LeSueur2008,Pillet2010,Chang2013,Pillet2013,Bretheau2013,Bretheau2013b,Bretheau2017}, the way ABS form in graphene coupled to large critical field superconductors and subject to high magnetic field remains unclear.
For this purpose, we have performed tunneling spectroscopy measurements of graphene proximitized by niobium/niobium-nitride (Nb/NbN) electrodes. The measured energy spectra reveal a strong proximity effect in graphene and the presence of a continuum of ABS, which varies with the graphene carrier density.
We also observe phase-dependent out-of-gap energy features associated with microscopic quantum dots, whose energy levels are coupled to both the proximitized graphene and the tunneling probe, and which indicate the presence of unintentional resonant impurities even in clean graphene-based van der Waals heterostructures.

\begin{figure*}
\includegraphics[width=2\columnwidth]{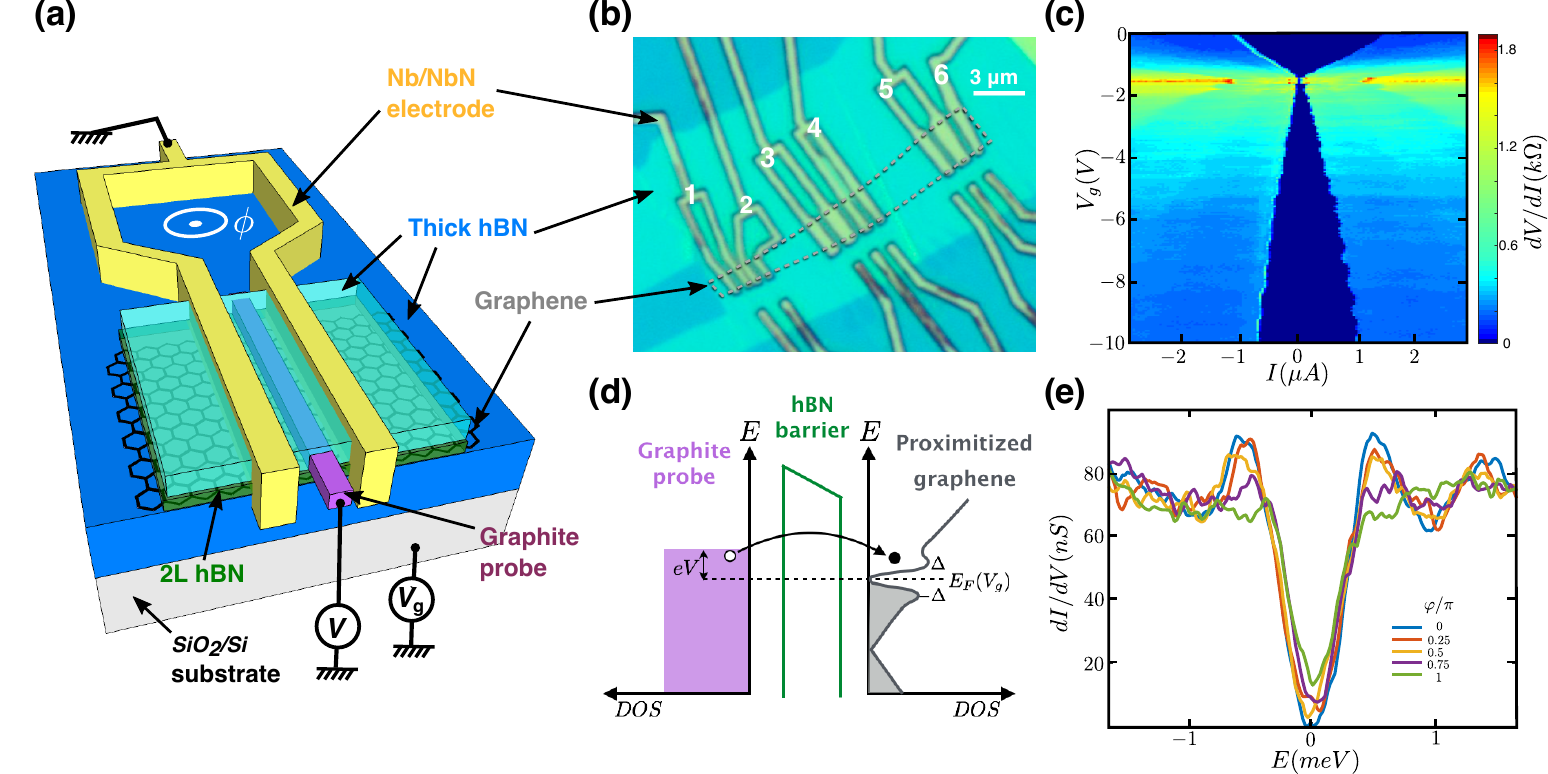}
\caption{
(a) Device structure. An encapsulated graphene flake is connected to two Nb/NbN superconducting electrodes. Magnetic flux $\phi$ threading the loop imposes a phase difference $\varphi=\phi / \phi_0$ across graphene.
(b) Optical micrograph of the sample showing several S-G-S devices, built on the same graphene flake. (c) Color-coded differential resistance, $dV/dI$, between junction \#5 and \#6 as a function of current bias, $I$, and gate voltage, $V_g$. The critical current exhibits Fabry-Perot oscillations for hole doping. (d) Schematics of the tunneling spectroscopy process. The normal probe is a graphite electrode and the tunneling barrier a bilayer hBN crystal. (e) Differential conductance of junction \#4, $dI/dV$, as a function of energy, $E=eV$, for different
phases, $\varphi$, and at a gate voltage $V_g=40~V$. The DOS exhibits a hard induced gap at zero phase.
\label{fig1}} \end{figure*}

The experiment is performed using a complex van der Waals heterostructure, schematized in Fig.~\ref{fig1}a, which is assembled with a polymer-based dry pick up and transfer technique~\cite{SupMat}. It consists of (from top to bottom) a hexagonal boron nitride (hBN) encapsulation layer, a microfabricated graphite probe, a bilayer-hBN tunneling barrier, a monolayer graphene sheet, and a bottom hBN substrate. The use of graphite as a probe reduces doping in graphene owing to their small work function mismatch while the top and bottom hBN flakes fully isolate graphene from contamination during the fabrication process. This strategy enables us to access the low carrier density regime of a pristine graphene sheet, whose Fermi energy is controlled electrostatically by the gate voltage $V_{g}$.
The encapsulated graphene (G) is connected at both ends via 1-D edge contacts~\cite{Wang2013} to two superconducting (S) electrodes made out of Nb/NbN, thus forming an S-G-S Josephson junction. The superconducting contacts are deposited onto the exposed graphene edges using e-beam evaporation of titanium as a sticking layer (5~nm) and \textit{in situ} reactive sputtering of Nb (15~nm) and NbN (50~nm). The superconductor prepared this way has a critical temperature $T_{c}\sim$ 9~K and remains superconducting at a magnetic field $B=9~$T (characterized separately), well above the onset of integer quantum Hall effect in a high quality graphene device. To control the superconducting phase difference $\varphi=\phi/\phi_0$ across the S-G-S Josephson junction, we apply a magnetic flux $\phi$ through the superconducting leads patterned in a loop geometry, where $\phi_0=\hbar/2e$ is the reduced flux quantum.

The full device actually consists of five superconducting loops built on the same monolayer graphene sheet (see Fig.~\ref{fig1}b). In each loop, the lead-to-lead distance $L$ for the graphene weak link is 440~nm, and the width $W$ ranges from 1.3~$\mu$m to 3.4~$\mu$m. The tunneling measurements presented in this paper are obtained from tunnel junction \#1 ($W=1.3$~$\mu$m) and tunnel junction \#4 ($W=2.7$~$\mu$m). Such a geometry allows for both spectroscopic and transport measurements in the same graphene flake. Indeed, by measuring at low temperature (20 mK) the current between two neighboring loops, one can extract the Josephson critical current through graphene. Such a measurement is shown in Fig.~\ref{fig1}c, as a function of the gate voltage $V_g$. The Fabry-Perot oscillations in the critical current observed in the hole (or $p$-doped) region ($V_g < -1.8$~V) demonstrate that transport is ballistic in the graphene junctions~\cite{BenShalom2016,Calado2015,Allen2016}.

The density of states (DOS) of graphene can be extracted by measuring (at 20 mK) the differential conductance $dI/dV$ between the graphite probe and the superconducting lead (Fig.~\ref{fig1}d). Figure~\ref{fig1}e shows such a measurement for junction \#4 at $V_{g}=40$~V, with $dI/dV$ plotted as a function of bias voltage $V$ (converted into energy $E=eV$) for different values of the magnetic flux $\phi$ (converted into phase $\varphi$). Due to the proximity effect, the extracted graphene DOS displays an induced gap $\Delta \sim$ 0.6 meV. It's 40$\%$ smaller than the gap size of NbN, which might be related to the antiproximity effect associated with the intermediate layers of Nb ($T_{c}\sim 7$~K) and Ti ($T_{c}\sim 0.4$~K). At $\varphi =0$, the DOS exhibits a hard induced gap (zero DOS at low energy), which demonstrates the strength of the proximity effect and the high transparency of the SG interfaces. The DOS oscillates when varying the phase, which reveals the presence of a continuum of ABS in graphene, as expected from a 2-D quantum conductor that accommodates a large number of conduction channels and associated ABS~\cite{Kulik1970,Furusaki1991,Beenakker1991,Bagwell1992}.

%
%

\begin{figure}[htbp]
\centering{}\includegraphics[width=1\columnwidth]{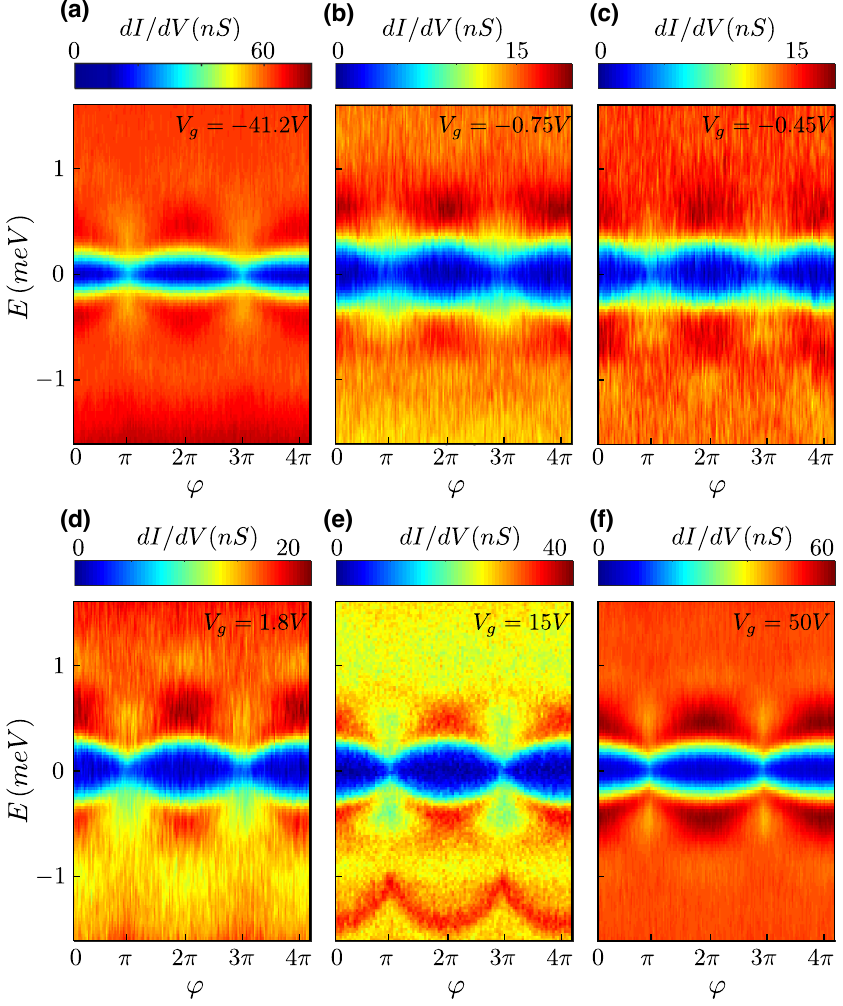}
\caption{
(a-f) Differential conductance of junction \#1, $dI/dV$, as a function of both energy $E=eV$ and superconducting phase difference $\varphi$, for different gate voltages $V_g$ (indicated in each panel).
The oscillating spectrum is evidence for a continuum of ABS.
\label{fig2}} \end{figure}

To investigate further how the proximity effect develops in graphene, we now tune the Fermi energy 
by varying the gate voltage $V_g$. 
The measured energy spectra, plotted in Fig.~\ref{fig2} as a function of both energy and phase for junction \#1, strongly depend on the carrier density, which is tuned from hole type (Fig.~\ref{fig2}a), through the charge neutrality point (CNP) (Fig.~\ref{fig2}b-\ref{fig2}d) to electron type (Fig.~\ref{fig2}e-\ref{fig2}f).
The phase modulation of the DOS is weaker when the graphene is hole-doped, in good agreement with transport measurements from Fig.~\ref{fig1}c that shows a big asymmetry in supercurrent between electron and hole-doped regions. This is due to the presence of $p$-$n$ junctions at the SG interfaces (owing to the $n$-type electron doping of the graphene contact region by the Nb/NbN electrodes), which reduce the contacts' transparency and weaken the phase modulation of the ABS.
Strikingly, the phase modulation remains well pronounced at the CNP,
which demonstrates the low disorder of our graphene nanodevice. This observation and the one of the hard gap are in clear contrast with measurements obtained in previous work using aluminum as a superconductor ~\cite{Bretheau2017}. Such improvements, combined with the use of large  critical field superconductors, are very promising for projects that require quantum coherence in clean hybrid superconducting Dirac materials~\cite{Lindner2012,Clarke2013,Mong2014,San-Jose2015}.


Going further, we explore Josephson effect at larger magnetic fields, both through transport and spectroscopic measurements (see Supplemental Material~\cite{SupMat}).
When a significant flux is threading graphene, the supercurrent is reduced (see Fig.~S3b), though not as quickly as expected. A striking departure from the conventional Fraunhofer pattern is indeed observed, as already reported in Refs.~~\cite{BenShalom2016,Calado2015}. This observation is consistent with tunneling spectroscopy measurements (see Fig.~S3a), which show that the DOS oscillations persist up to 15~mT. This large-field Josephson effect might be related to nontrivial ABS that persist at the graphene edge.

\begin{figure*}
\includegraphics[width=2\columnwidth]{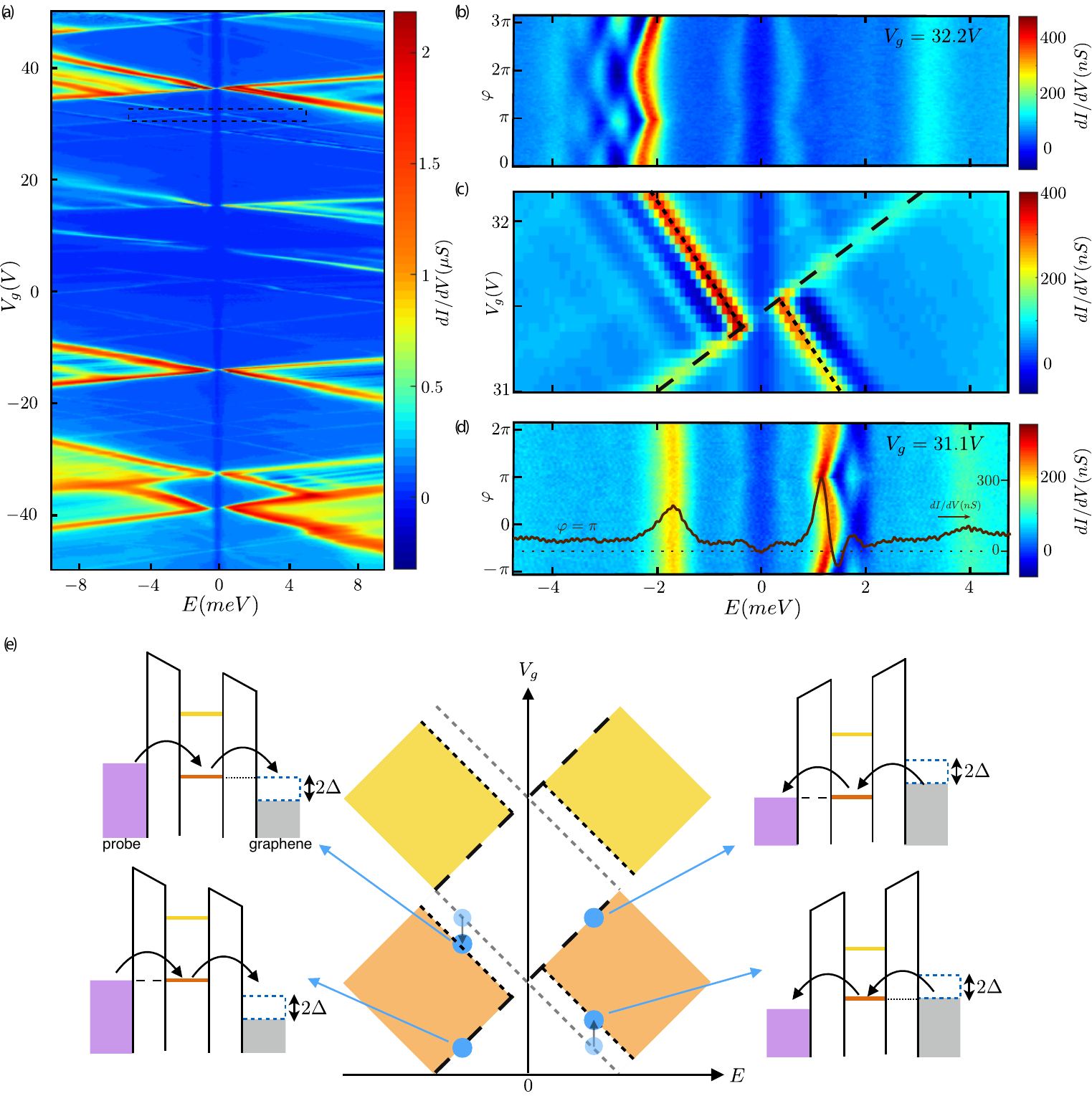}
\caption{
(a) Differential conductance of junction \#4, $dI/dV$, as a function of both energy $E=eV$ and gate voltage $V_g$. (c) Zoom-in at the crossing of one Coulomb diamond, highlighted by a dashed rectangle in (a). The dashed lines highlight the splitting of the negative slope diamond peaks (NSDP), while the positive slope diamond peaks (PSDP) remain aligned across the induced gap. (b), (d) Phase dependence of NSDP at $V_g=32.2$~V and $31.1$~V, respectively. The $dI/dV$ linecut at $\varphi$ = $\pi$ in (d) shows the change of sign, which is associated with the graphene proximitized DOS.
(e) Middle: schematics of the Coulomb diamonds with gap opening. Around: schematics of transport through a quantum dot connected to a normal and a superconducting electrode. 
\label{fig3}} \end{figure*}

This set of measurements establishes that a continuum of supercurrent-carrying ABS form in graphene, with energies $|E|<\Delta$. At some specific gate voltages however (see, e.g., the oscillating red feature for $E < -1$ meV in Fig.~\ref{fig2}e), we observe resonant out-of-gap energy features that depend on the phase difference. To understand their origin, we measure the differential conductance as a function of both energy and gate voltage, at a constant phase and over a large energy range. As shown in Fig.~\ref{fig3}a,  on top of the induced gap that appears at low energy, one can see sharp resonances that disperse in energy and gate voltage, reminiscent of Coulomb diamonds.
A detailed analysis suggests that they correspond to 5-20 nm size quantum dots (QD) with typical addition energy of $\sim$ 5-60 meV.
These quantum dots might be related either to  spurious defects embedded in the van der Waals heterostructure
or to charge puddles in graphene around charged impurities~\cite{Amet2012}. Another, more intrinsic, explanation could be scattering centers at the graphene edge, which were recently evidenced as being located every 2-20~nm using scanning nano-SQUID thermometry~\cite{Halbertal2017}.

Figure.~\ref{fig3}c shows a zoom-in on a given diamond at low energy. Strikingly, the diamond boundary peaks in $dI/dV$ that disperse in $(eV,V_g)$ with a negative slope (hereafter called NSDP) split around zero energy, while the positive slope diamond peaks (PSDP) are aligned. Moreover, the NSDP are accompanied by $dI/dV$ peaks of opposite sign (see inset in Fig.~\ref{fig3}d). Similar effects were already observed in S-QD-S hybrid systems, using metallic nanoparticles~\cite{Ralph1995,Black1996}, carbon nanotubes~\cite{Eichler2007}, semiconductor nanowires~\cite{Doh2008}, or fullerene molecules~\cite{Winkelmann2009}. However, the configuration here is asymmetric with a QD weakly coupled on one side to the graphite probe and on the other side to the proximitized graphene. The NSDP (respectively, PSDP) thus correspond to the alignment of the resonant dot level with the peak at the edge of the induced superconducting gap in the graphene DOS  (resp. with the Fermi level of the graphite probe), as schematized in Fig.~\ref{fig3}e. Further, when the phase difference is varied, the graphene DOS is modulated and the NSDP oscillate in energy. This phenomenon can therefore happen at positive or negative energy, depending on the gate voltage (Fig.~\ref{fig3}b and \ref{fig3}d).

In this experiment, the quantum dots behave as energy filters in the tunneling process from graphite to graphene, with a rate that depends on both energy and gate voltage. Out of resonance, the tunneling rate is weak, it hardly depends on energy, and one directly probes the graphene DOS by measuring the differential conductance. At resonance, the differential conductance is greatly increased and the tunneling rate strongly depends on energy. In the simple case of a very weakly coupled QD, the tunneling rate can be modeled as a Dirac delta function. The $dI/dV$ signal is then proportional to the derivative of the graphene DOS, which explains the observed positive-followed-by-negative $dI/dV$ values.



In conclusion, using tunneling spectroscopy in a full van der Waals heterostructure, we demonstrated  that graphene coupled to Nb/NbN superconductors can develop a strong proximity effect, with a DOS displaying a hard induced gap and a pronounced phase modulation near the charge neutrality point. These results open the way to exploring exotic Andreev physics at large magnetic field, using ultraclean Dirac materials coupled to large  critical field superconductors ~\cite{Lindner2012,Clarke2013,Mong2014,San-Jose2015}.
Furthermore, our measurements
reveal the presence of microscopic quantum dots weakly coupled to the proximitized graphene, that behave as energy filters in the tunneling process. To elucidate their origin, which could be intrinsic and associated to the graphene finite dimensions, one could combine tunneling spectroscopy and nano-SQUID thermometry measurements.

\begin{acknowledgements}
We thank Amir Yacoby, Sean Hart, and Di Wei for the support in fabricating superconducting (Ti/Nb/NbN) electrodes.
This work has been primarily supported by the U.S. DOE, BES Office, Division of Materials Sciences and Engineering under Award DE-SC0001819, by the Gordon and Betty Moore Foundation's EPiQS Initiative through Grant No. GBMF4541 to PJH and by Obra Social "la Caixa" Fellowship to DRL. This work made use of the MRSEC Shared Experimental Facilities at MIT, supported by the National Science Foundation under award number DMR-14-19807 and of Harvard CNS, supported by NSF ECCS under award No. 1541959.
\end{acknowledgements}


\end{document}